\documentclass[aps,prl,amsmath,amssymb,superscriptaddress,nofootinbib,twocolumn,floatfix]{revtex4-1}
\usepackage{graphicx}
\usepackage[usenames]{color}
\usepackage[colorlinks=true,linkcolor=cyan,citecolor=cyan,anchorcolor=green,urlcolor=magenta]{hyperref}
\usepackage{subfigure}

\def\be{\begin{equation}}
\def\ee{\end{equation}}
\def\ba{\begin{aligned}}
\def\ea{\end{aligned}}

\begin{document}

\title{A Proposal for Detecting Superfluidity in Neutron Stars}

\author{Yunjing Gao}
\affiliation{Tsung-Dao Lee Institute, Shanghai Jiao Tong University, Shanghai 201210, China}

\author{Jiahao Yang}
\affiliation{Tsung-Dao Lee Institute, Shanghai Jiao Tong University, Shanghai 201210, China}

\author{Zhenyu Zhu}
\affiliation{Tsung-Dao Lee Institute, Shanghai Jiao Tong University, Shanghai 201210, China}

\author{Yosuke Mizuno}
\email{mizuno@sjtu.edu.cn}
\affiliation{Tsung-Dao Lee Institute, Shanghai Jiao Tong University, Shanghai 201210, China}
\affiliation{School of Physics and Astronomy, Shanghai Jiao Tong University, Shanghai 200240, China}

\author{Jianda Wu}
\email{wujd@sjtu.edu.cn}
\affiliation{Tsung-Dao Lee Institute, Shanghai Jiao Tong University, Shanghai 201210, China}
\affiliation{School of Physics and Astronomy, Shanghai Jiao Tong University, Shanghai 200240, China}

\date{\today}

\begin{abstract}
Based on the GW dispersion relation raised in \cite{Cetoli2012},
we investigate the possible reflection of gravitational wave (GW) by superfluidity (SF) in the neutron star,
provided its high density and dissipationless properties.
Following this scenario,
an experimental proposal is raised to probe the expected
SF in neutron star
by means of GW detection.
Two types of binary systems are considered, neutron star-black hole and binary neutron star systems,
with weak gravitational field condition imposed.
Non-negligible modulation on the total signal caused by the GW reflection is found, which contributes amplitude and phase variations distinguishable from the primitive sine signal.
Furthermore,
we show that it is possible for such modulations to be detected
by the Cosmic Explorer and Einstein Telescope at $100\,\mbox{Mpc}$.
Identification of those signals can
evince the existence of
the long-sought SF in
neutron stars as well as the exotic superfluidity-induced GW reflection.

\end{abstract}

\maketitle

\paragraph*{Introduction.---}
\label{sec:intro}
Superfluidity (SF) in neutron star (NS)
has been predicted and discussed for a long time \cite{migdal_superfluidity_1959,baym_electrical_1969}.
Given high stellar density in NS, roughly $10^{14}g/cm^3$,
the neutron pairing is expected to take place in the crust as well as outer core since its critical temperature ($\sim10^{10}K$)
is much larger than the typical temperature of NSs except for the formation stage \cite{ginzburg_superfluidity_nodate,migdal_superfluidity_1959,Wolf}.
In addition,
for the charged particles,
superconductivity of protons is possible at the
crust-core interface,
with the critical temperature on the same scale as neutron pairing \cite{protonSC}.
In contrast,
pairing of compressed electrons in NS is excluded,
due to its exponentially small critical temperature
$T_c\approx3\times10^3\rho_e^{1/6}e^{-2.8\rho_e^{1/3}}K$
with the electron density $\rho_e\sim10^{12}g/cm^3$
\cite{ginzburg_superfluidity_nodate}.
Continuous efforts have been devoted to verifying SF in NS.
One possible evidence comes from the observation of glitch (e.g., \cite{pulsar}),
and a widely accepted explanation attributes such phenomenon to release of pinned vorticity of
the neutron SF, which leads to angular momentum transfer in NS \cite{Anderson}.
From another point of view,
Ref.~\cite{heinke_direct_2010} shows that
the cooling procedure of the Cassiopeia A Neutron Star is 4\% faster than
conventional prediction \cite{cooling_prediction} during the last ten years,
whose cooling mechanism can be explained
by enhanced neutrino emission during pairing formation of neutrons \cite{page_rapid_2011}.

The observations of GW170817 \cite{BNS_LIGO}, GW200105 and GW200115 \cite{NSBH_LIGO} capture the emission of GW in a binary neutron star (BNS) system and neutron star-black hole (NSBH) systems,
providing a possibility to connect the detection of SF in NS to GW observation.
From the GW dispersion relation raised in the study of GW interaction with fermionic SF \cite{Cetoli2012},
the contribution can be divided into ``diamagnetic'' and ``paramagnetic'' parts.
A similiar mechanism has also been pointed out in \cite{Gratz}.
The absorption for GW is described by the imaginary part of the paramagnetic term,
which corresponses to the quasiparicle excitation in the SF discussed here.
Competition between the ``diamagnetic'' and ``paramagnetic'' terms can lead to different propagation scenarios of GW,
including a regime where the GW will be reflected.

If the reflection from NS can happen,
it will give rise to a modulation for the observed GW signals.
In return, a direct observation of the modulation can provide
one-stone-two-bird evidence: 1, it supports the existence of SF in NS;
2, it implies the GW reflection from SF.
In this letter,
first we show an estimation on the possible GW reflection by SF in the NS,
corresponding to typical GW frequency measurable nowadays.
Based on this feature, we study the modulation signal caused by reflection from two systems: the NSBH system and the BNS system.
The modulation signal contributes non-negligible amplitude and phase deviations,
whose variation with GW frequency and mass of the NS (or black hole) is studied.
Furthermore, we show that the modulation can possibly be probed by
the Cosmic Explorer as well as Einstein Telescope in a reasonable parameter region.

\paragraph*{Mechanism for GW reflection.---}
\label{sec:mechanism}

The response of superfluidity to incident GW can be described by the GW dispersion relation \cite{Cetoli2012}
\be
\omega^2_{\rm GW}\approx c^2k_{\rm GW}^2+\frac{32\pi GP}{c^2}+\frac{16\pi G\chi}{c^2},
\ee
where $P$ stands for the pressure in SF and $\chi$ is the response function of stress-energy tensor variation to the metric perturbation.
The second term on the right hand side serves as a ``diamagnetic'' term to the GW,
which comes from the metric dependency of stress-energy tensor.
The third term referred to as ``paramagnetic'' term has a non-local nature,
which originates from the variation of SF density matrix with respect to the metric \cite{Gratz},
and can be described by the linear response theory \cite{linear}.
Since the energy scale of typical NS temperature is much smaller than the pairing gap,
the results derived by the zero temperature formalism is a reasonable approximation for the cases here.

Given the degeneracy pressure about $1.6 \times 10^{34} \mbox{ Pa}$ in the NSs \cite{pressure},
the diamagnetic part dominates the dispersion.
Simple calculation tells us that
when $\omega_{\rm GW} \lesssim \omega_{cut} \approx 3\times10^4 \mbox{ Hz}$
the GW has a purely imaginary wave vector
resulting in a finite penetration depth correspondingly.
Moreover, for typical GW, as observed in real detections,
its frequency is much smaller than $\omega_{cut}$, as such
the penetrati
on depth is approximately $8.6\times10^3 \mbox{ m}$ which is
a few times smaller than the NS diameter.
Furthermore,
within the zero temperature approximation,
$\rm Im[\chi]=0$ as typical GW frequency $\omega_{\rm GW} \ll 2\Delta/\hbar$ ($\Delta$ is the pairing gap).
The vanishing imaginary part indicates no absorption of GW energy by the SF,
combined with the finite penetration depth,
significant reflection of GW is expected in the dense and dissipationless superfluid layer of NSs.

\paragraph*{The NSBH system.---}
\label{sec:NSBH}

\begin{figure}[t]
\centering
\includegraphics[width=0.75\columnwidth]{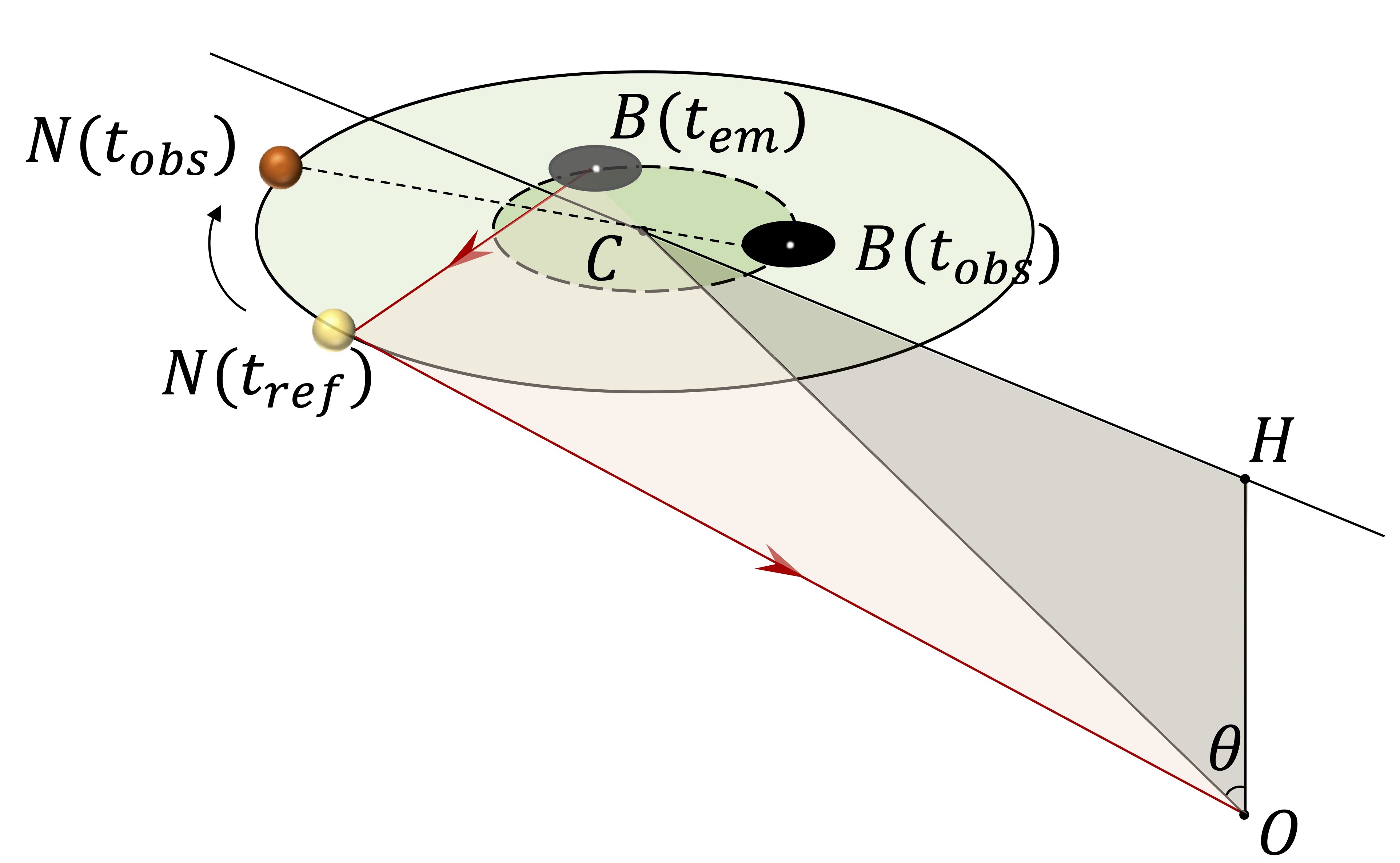}
\caption{Illustration for GW reflection in the NSBH system.
Here $N$, $B$, $C$ and $O$ stand for NS, BH, rotation center and observer,
respectively.
$\theta$ denotes the inclination angular of observer.
Plane $CHO$ is perpendicular to the orbital plane which crosses the perigee.
The red lines and arrows track the rerflection signal propagation.
In the local time of observer,
the orange-shaded ball illustrates
real-time position for the NS,
and the reflection signal received at this time comes from $N(t_{ref})$ (yellow-shaded ball) at some earlier time,
which is emitted at $B(t_{em})$ by the BH.
}
\label{fig:outplane_case_of_NSBH}
\end{figure}

The total signal ($E_{tot}$) consists of direct incident ($E_{dir}$) and reflection ($E_{ref}$) parts.
For $E_{dir}$,
quadrupole moment approximation is applied to determine the amplitude, which considers the NS and black hole (BH) as a whole contributor.
Both NS and BH are assumed to rotate around the mass center $C$ in a circular motion without eccentricity.
The propagation distance before observation is taken as $|CO|$,
and $|\cdot|$ stands for the distance between two points.
By taking Newtonian orbital approximation, the orbital radius $R_{orb}=\left(GM_{\rm{BH}}/\omega_{\alpha}^2\right)^{1/3}$,
where the angular velocity $\omega_{\alpha} = \omega_{\rm GW}/2$, and $G$ is the gravitational constant.
For direct signal received at $t_{obs}$
\be
E_{dir}(t_{obs})=\frac{\mathcal{E}_0^{dir}}{|CO|}e^{-i\phi_0(t_{obs})},
\label{eq:E_dir_NSBH}
\ee
where $\mathcal{E}^{dir}_0 = 4 G R_{orb}^2 M_{\rm{NS}} \omega_{\alpha}^2/c^4$,
$\phi_0(t)=\omega_{\rm{GW}}t-k_{\rm{GW}}|CO|$
with $k_{\rm{GW}}$ being the GW wave vector and $c$ being the speed of light.

For the reflection part,
the GW on the surface of NS is assumed to be fully reflected for simplicity, as the NS radius is larger than the penetration depth.
Due to the complexity of GW emission from the source and the reflection mechanism, a phenomenological model is employed to simplify this process.
Contribution from BH to the GW is extracted as a result of quadrupole moment variation of BH rotating around the center of mass (C in Fig.~\ref{fig:outplane_case_of_NSBH}).
This part of GW propagates outward spherically from $B(t_{em})$ and is reflected on the surface of NS at a later time $t_{ref}$ with amplitude $\mathcal{E}_0^{ref}/|B(t_{em})N(t_{ref})| = 4G R_{\rm BH}^2 M_{\rm BH} \omega_{\alpha}^2/(c^4|B(t_{em})N(t_{ref})|)$, where $R_{(\cdot)}$ stands for orbital radius.
The NS plays the role of ``spherical  mirror'' of GW.
Since the typical radius of NSs ($r_{\rm{NS}}\approx 20\, \text{km}$) is much smaller than the wavelength of GW of interest,
isotropic reflection is considered.
After being scattered by the NS,
amplitude of the outgoing GW which propagates to the observer $O$ at time $t_{obs}$ becomes $(\mathcal{E}^{ref}_0/|ON(t_{ref})|)\cdot (r_{\rm{NS}}/|B(t_{em})N(t_{ref})|)$.

Distances between NS and BH located at different times have been encountered here,
which can be determined by finding the propagation time relation.
Both $t_{em}$ and $t_{ref}$ are referred to as functions of $t_{obs}$,
whose relations are shown in \cite{NSBH_app}.
In the following,
we denote the propagation distance before and after reflection corresponding
to GW observed at $t_{obs}$ as $L_b(t_{obs}) = c(t_{ref}-t_{em})$ and $L_a(t_{obs}) = c(t_{obs}-t_{ref})$.
As a result,
the reflection signal can be formally given by
\be
E_{ref}(t_{obs}) = \frac{\mathcal{E}^{ref}_0r_{\rm{NS}}e^{-i[\omega_{\rm GW}t_{obs}-k_{\rm GW}(L_a(t_{obs})+L_b(t_{obs}))]}}{L_a(t_{obs})L_b(t_{obs})}
\label{eq:E_ref_NSBH}
\ee

\begin{figure}[b]
    \centering
    \includegraphics[width=0.5\textwidth]{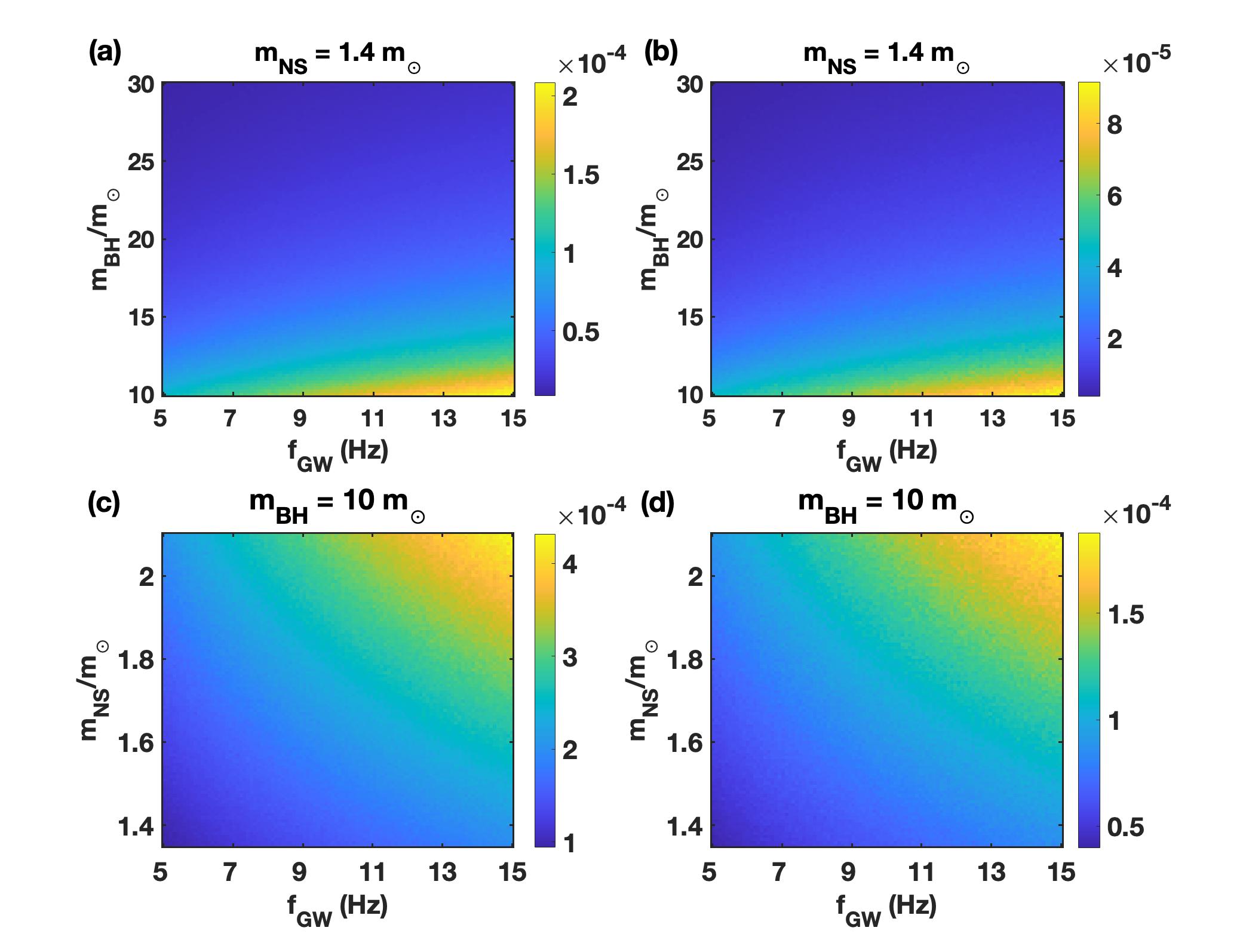}
    \caption{Plot of $\sigma_{\mathcal{A}}$ in (a, c) and $\sigma_{\phi}$ in (b, d) for the NSBH system with $\theta=\pi/2$. The first row is fixed for $m_{\rm NS}=1.4\,m_{\odot}$ and the second row with fixed $m_{\rm BH} = 10 \,m_{\odot}$.}
    \label{fig:fluctuation_NSBH}
\end{figure}

\begin{figure}[t]
\centering
\includegraphics[width=0.7\columnwidth]{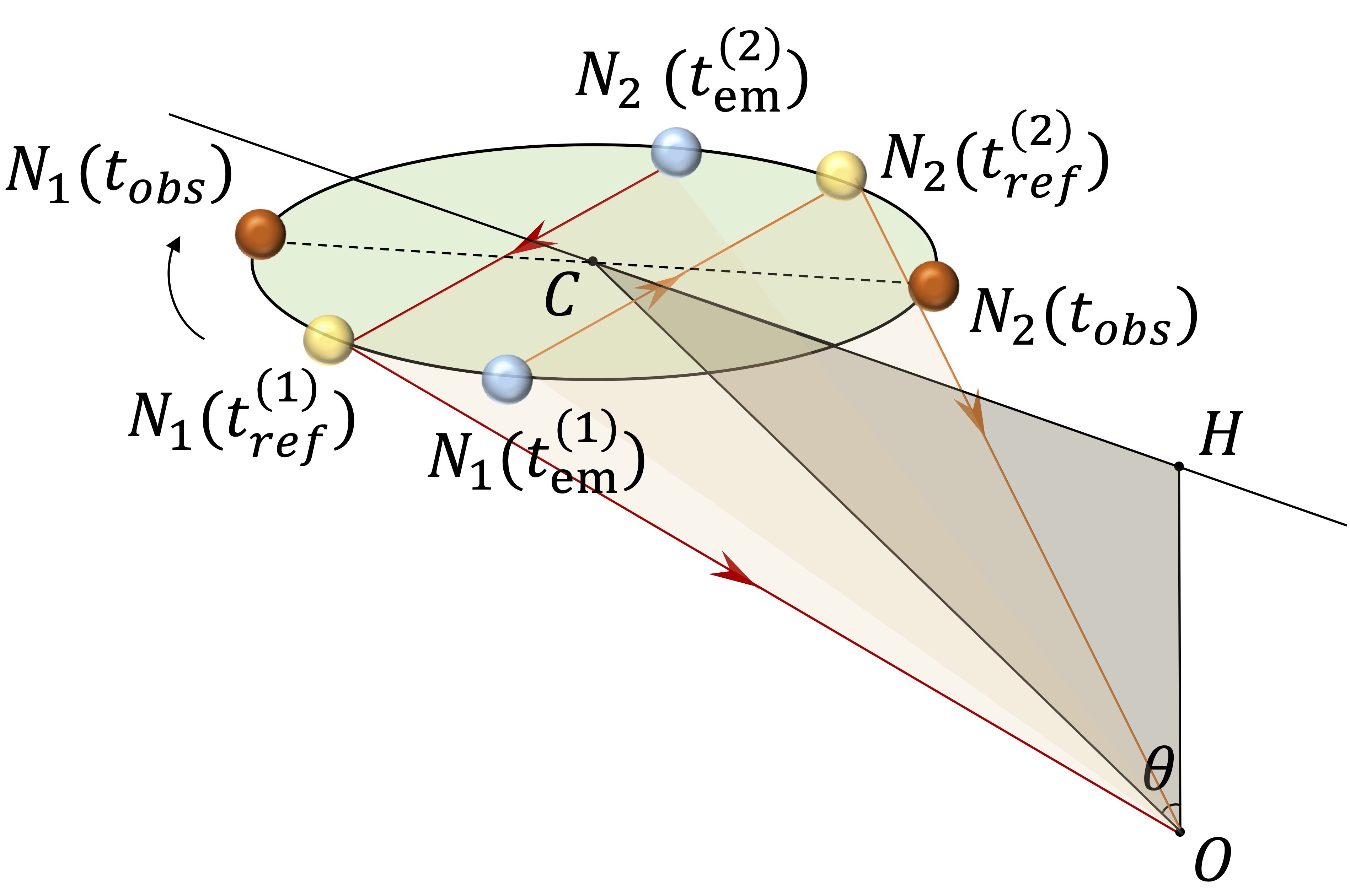}
\caption{ Illustration for GW reflection in the BNS system.
$C, H, O,\,\theta$, ibid.(Fig. \ref{fig:outplane_case_of_NSBH}).
The orange and red arrows and lines show the paths of reflection signals.
Two orange-shaded balls label the real-time positions of the two NSs.
Yellow-shaded balls are the positions of NSs
when reflection happens,
each related to one blue-shaded ball denoting the emission position. }
\label{fig:Outplane_BNS}
\end{figure}

As a simple example,
in the limit $k(R_{\rm BH}+R_{\rm NS})\ll 1$,
$t_{em}$ can be treated as a constant shift of $t_{ref}$.
Also,
since $|CO|\gg R_{\rm BH}$,
the reflection signal can be simplified as
\be \ba
&E_{ref}^{lim}(t_{\rm obs}) = \frac{\mathcal{E}_0^{ref}}{|CO|}\frac{r_{\rm NS}e^{-i[\phi_0(t_{obs})+\kappa \cos[\omega_{\alpha}(t_{obs}-|CO|/c)]]}}{R_{\rm BH}+R_{\rm NS}},
\label{eq:NSBH_ref_simplify}
\ea \ee
with $\kappa  = k_{\rm GW}R_{\rm NS}\sin\theta$.
The contribution of reflection signal can be seen clearly if we consider the Fourier component of Eq.\eqref{eq:NSBH_ref_simplify}.
Apart from the dominant contribution at $f_{\rm{GW}} = \omega_{\rm GW}/2\pi$,
additional components at $f_{\rm{GW}}/2$, $3\,f_{\rm{GW}}/2$, $2\,f_{\rm{GW}}$, $\cdots$ show up.
This can be viewed from the expansion of $\mbox{exp}[i\kappa\cos(\omega_{\alpha}t_{obs})]$,
where $\kappa^n\cos^n(\omega_{\alpha}t_{obs})$ can be rewritten as a summation
of $\kappa^n\mbox{exp}[\pm im(\omega_{\alpha}t_{obs})]$ terms with $m,\,n\in N$.
The additional peaks lead to distinct difference in the observation result.

In the general case,
$E_{tot}$ can be expressed in a compact form after combining Eq.~\eqref{eq:E_dir_NSBH} and Eq.~\eqref{eq:E_ref_NSBH},
and we define $\gamma(t) = (\mathcal{E}_0^{ref}/\mathcal{E}_0^{dir})(r_{NS}/L_b(t))$.
Since it is sufficient to consider the properties of one GW component,
then
\be
\begin{aligned}
{\rm Re}[E_{tot}(t_{obs})]=\frac{\mathcal{E}_0}{|CO|}\mathcal{A}(t_{obs})\cos\left[\phi_0(t_{obs})-\phi(t_{obs})\right],
\end{aligned}
\label{eq:real_E_total_NSBH}
\ee
where
\be
\begin{aligned}
&\mathcal{A}(t)=\sqrt{1+\gamma^2(t)+2\gamma(t)\cos[k_{\rm{GW}}(L_b(t)+L_a(t)-|CO|)]}\\
&\tan\phi(t)=\frac{\gamma(t)\sin\left[k_{\rm{GW}}(L_b(t)+L_a(t)-|CO|)\right]}{1+\gamma(t)\cos\left[k_{\rm{GW}}(L_b(t)+L_a(t)-|CO|)\right]}.
\end{aligned}
\label{eq:A_phi}
\ee

As a consequence of $E_{ref}$,
the explicit modulations appear in
the relative amplitude $\mathcal{A}(t)$ and the non-linear phase $\phi(t)$ of $\mbox{Re}[E_{tot}]$.
Both modulations are
at the same order of magnitude of $\gamma$,
whose effect can be viewed from the standard deviations of the two quantities,
\be
\begin{aligned}
&\sigma_{\mathcal{A}}=\sqrt{\langle\mathcal{A}^2\rangle-\langle\mathcal{A}\rangle^2}\\
&\sigma_{\phi}=\sqrt{\langle\phi^2\rangle-\langle\phi\rangle^2},\\
\end{aligned}
\label{eq:fluctuation_NSBH}
\ee
where $\langle\cdot\rangle=\int_0^{T_{orb}}\cdot\, dt/T_{orb}$ with $T_{orb}$
being the orbital period.
As can be observed from Eq.~\eqref{eq:A_phi},
more significant fluctuations occur for larger $\gamma$ and $k_{\rm GW}$.
Moreover, $\gamma\propto M_{\rm NS}M_{\rm BH}^{1/3}/(M_{\rm BH}+M_{\rm NS})^{5/3}$ approximately.
This is consistent with the result of $\sigma_A$ and $\sigma_{\phi}$ shown in Fig.~\ref{fig:fluctuation_NSBH}.
Larger $M_{\rm NS}$ leads to stronger fluctuations for both $\sigma_A$ and $\sigma_{\phi}$,
while it is opposite for $M_{\rm BH}$.
As $f_{\rm GW}$ gets larger,
the fluctuations are significantly enhanced.

\begin{figure}[t]
    \centering
    \includegraphics[width=0.5\textwidth]{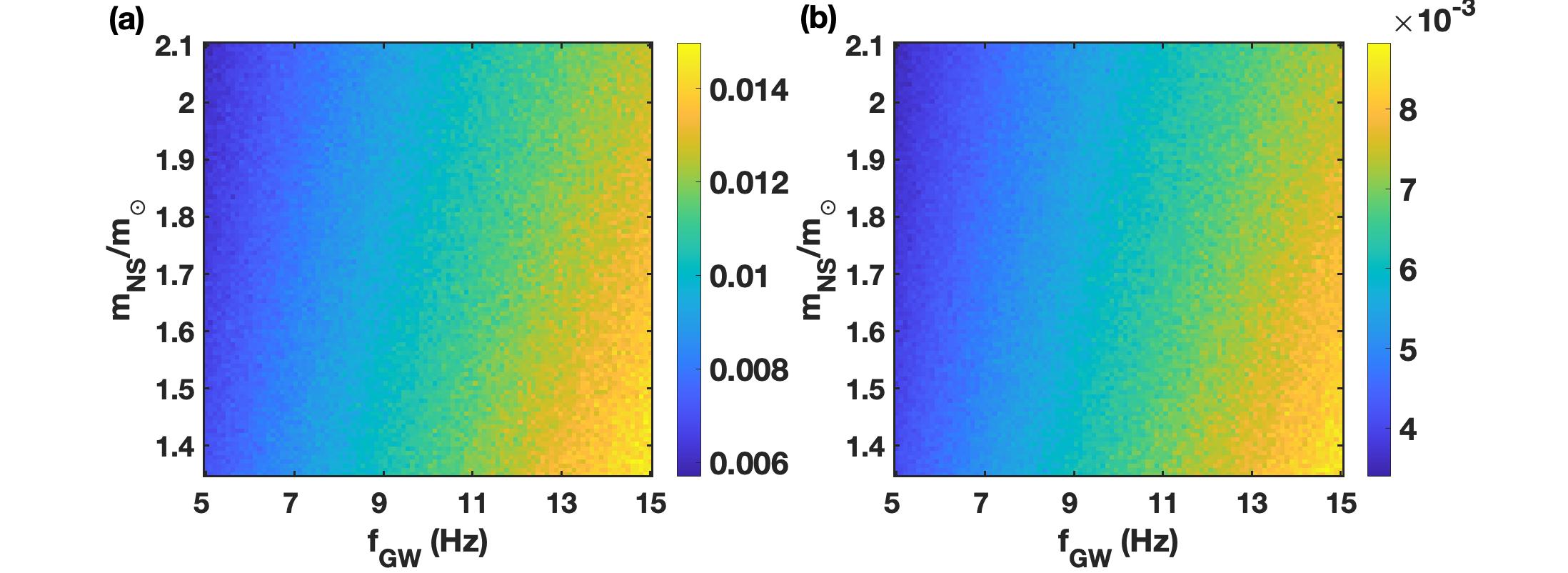}
    \caption{Plot for
     $\sigma_{\mathcal{\widetilde A}}$ (a) and $\sigma_{\widetilde\phi}$ (b) in the BNS system with $\theta=\pi/2$.}
    \label{fig:fluctuation_BNS}
\end{figure}

\paragraph*{The BNS system.---}
\label{sec:BNS}

As in the NSBH case
the same treatment is applied in this system.
Two NSs, $N_1$ and $N_2$ in Fig.~\ref{fig:Outplane_BNS}, are assumed to have equal mass.
Then the direct incident signal generated by the whole system take the form
\be
\widetilde{E}_{dir} = \frac{\widetilde{\mathcal{E}}_0^{dir}}{|CO|}e^{-i\widetilde{\phi}_0(t_{obs})},
\ee
with $\widetilde{\phi}_0(t_{obs}) = \omega_{\rm GW}t_{obs}-k_{\rm GW}|CO|$ and $\mathcal{E}_0^{dir} = 8\pi G \widetilde{R}_{NS}^2 M_{NS}\omega_{\alpha}^2/c^4$
with the orbital radius $\widetilde{R}_{NS}=(GM_{\rm{NS}}/4\omega_{\alpha}^2)^{1/3}$.
We use $\widetilde E_{ref}^{(\cdot)}$ and $L_{a(b)}^{(\cdot)}$ to represent reflection signal and propagation distance before ($b$) and after ($a$) the reflection happens,
which is related to reflection on $N_{(\cdot)}$.
Then the reflection signal is given by
\be
\begin{aligned}
&\widetilde E_{ref}(t_{obs})= \widetilde E_{ref}^{(1)}(t_{obs})+\widetilde E_{ref}^{(2)}(t_{obs})\\
=& \frac{\widetilde{\mathcal{E}}_0^{ref}}{L_a^{(1)}(t_{obs})}\frac{r_{NS}}{L_b^{(1)}(t_{obs})}
e^{-i[\omega_{\rm GW}t_{obs}-k_{\rm GW}(L_a^{(1)}(t_{obs})+L_b^{(1)}(t_{obs}))]}\\
+&\frac{\widetilde{\mathcal{E}}_0^{ref}}{L_a^{(2)}(t_{obs})}\frac{r_{NS}}{L_b^{(2)}(t_{obs})}
e^{-i[\omega_{\rm GW}t_{obs}-k_{\rm GW}(L_a^{(2)}(t_{obs})+L_b^{(2)}(t_{obs}))]},
\end{aligned}
\label{eq:E_ref_BNS}
\ee
where $\widetilde{\mathcal{E}}^{ref}_0 =  4 G \widetilde{R}_{orb}^2 M_{\rm{NS}} \omega_{\alpha}^2/c^4$ and the distance realtions can be found in  \cite{BNS_app}.
In the limit $k_{\rm GW}R_{\rm NS}\ll 1$,
$t_{ref}$ can be treated as a constant shift of $t_{obs}$,
leading to the approximation result
\be \ba
\widetilde{E}_{ref}^{lim}(t_{obs})
= \frac{\widetilde{\mathcal{E}}^{ref}_0}{|CO|}\frac{r_{\rm NS}}{|2R_{\rm NS}|}e^{i(\omega_{\rm GW}t_{obs}-2k_{\rm GW}R_{\rm NS})}\\
\times 2\cos\left(\kappa\cos[\omega_{\alpha}(t_{obs}-|CO|/c)]\right).
\ea \ee
Frequency components in addition to $f_{\rm GW}$ appear as can be observed by expanding  $\cos(\kappa\cos(\omega_{\alpha}(t_{obs})))$ in the small $\kappa$,
whose values take $0,\,2f_{\rm GW},\,3f_{\rm GW},\cdots$.

In the observation, the two branches of reflection signal superimpose on the direct incident
part, giving rise to the following result for real part of the total signal
\be
\begin{aligned}
{\rm Re}[\widetilde E_{tot}(t_{obs})]=\frac{\widetilde{\mathcal{E}}^{dir}_0}{|CO|}\mathcal{\widetilde A}(t_{obs})\cos[\widetilde{\phi}_0(t_{obs})-\widetilde\phi(t_{obs})],
\end{aligned}
\ee
where
\be
\begin{aligned}
&\mathcal{\widetilde A}(t)=\left(1+2\widetilde{\gamma}(t)(\cos\widetilde{\phi}^{(1)}+\cos\widetilde{\phi}^{(1)})\right.\\
&\left.+\widetilde{\gamma}(t)^2[2+2\sin\widetilde{\phi}^{(1)}\sin\widetilde{\phi}^{(2)}+2\cos\widetilde{\phi}^{(1)}\cos\widetilde{\phi}^{(2)}]\right)^{\frac{1}{2}}
\end{aligned}
\label{eq:A_bns}
\ee
and
\be
\tan\widetilde\phi(t)=\frac{\sin\widetilde{\phi}^{(1)}
+\sin\widetilde{\phi}^{(2)}}{1+\widetilde{\gamma}(t)(\cos\widetilde{\phi}^{(1)}+\cos\widetilde{\phi}^{(2)})},
\label{eq:phi_BNS}
\ee
with
\be \ba
&\widetilde{\phi}^{(1)} = \sin[k_{\rm{GW}}(L^{(1)}_a(t)+L_b^{(1)}(t)-|CO|)]\\
&\widetilde{\phi}^{(2)} = \sin[k_{\rm{GW}}(L^{(2)}_a(t)+L_b^{(2)}(t)-|CO|)].
\ea \ee

The relative amplitude $\mathcal{\widetilde A}(t)$ and non-linear phase  $\widetilde\phi(t)$ in $\mbox{Re}[E_{tot}]$ show explicit modulation in the presence of $\widetilde{E}_{ref}$.
Standard deviation $\sigma_{\mathcal{\widetilde A}}$ and $\sigma_{\widetilde\phi}$ can be defined in the same way as Eq.~\eqref{eq:fluctuation_NSBH}.
Also, the modulations in $\mathcal{\widetilde A}(t)$ and $\widetilde\phi(t)$ are
at the order of $\widetilde{\gamma}$,
which is approximately proportional to $M_{\rm NS}^{-1/3}$.
As a result,
both $\sigma_{\mathcal{\widetilde A}}$ and $\sigma_{\widetilde\phi}$ exhibit increasing
tendencies with larger $f_{\rm GW}$ but smaller $M_{\rm{NS}}$ as is shown in Fig.~\ref{fig:fluctuation_BNS}.
Those non-negligible fluctuations cause distinguishable modifications compared with the non-reflection model.

\begin{figure}[t]
    \centering
    \includegraphics[width=0.45\textwidth]{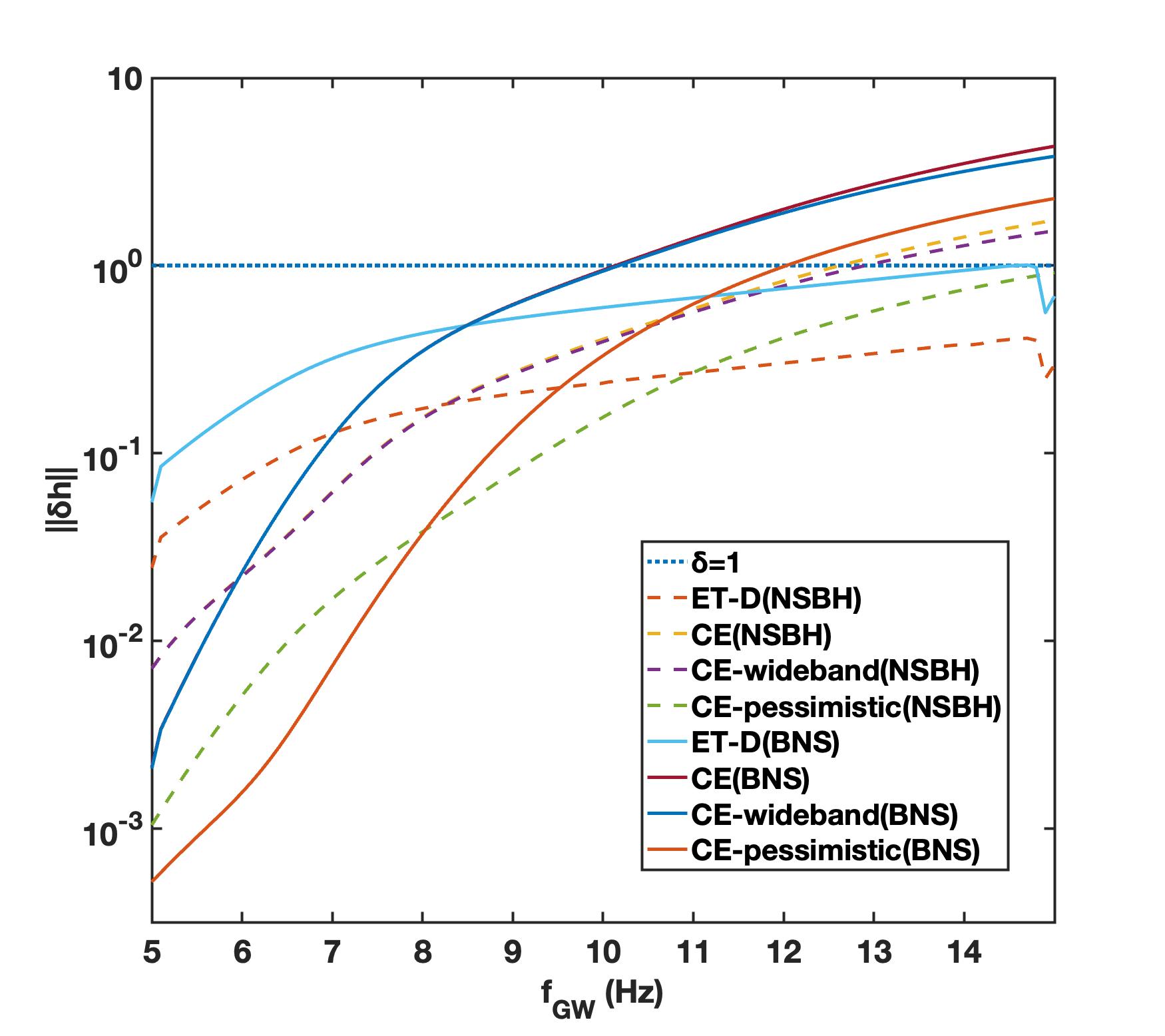}
    \caption{Plot of $||\delta h||$ at $100\,\mbox{Mpc}$ with the Cosmic Explorer and Einstein Telescope for NSBH and BNS systems.
    The bold blue line plots $||\delta h|| = 1$ as a reference.}
    \label{fig:SNR}
\end{figure}

\paragraph*{Signal distinguishability.---}
In this section we estimate the practical ability for the modulated signal caused by GW reflection to be distinguishable in GW detection.
As the discussion in \cite{SNR}, two waveforms $h_1$ and $h_2$ are distinguishable in principle if
\be
\begin{aligned}
||\delta h|| \equiv \sqrt{\langle h_1-h_2|h_1-h_2 \rangle}\gtrsim1
\end{aligned}
\ee
is satisfied,
where the noise-weighted inner product is given by
\be
\begin{aligned}
\langle h_1-h_2|h_1-h_2 \rangle =
4\,\mbox{Re}\int_0^{\infty} \frac{|h_1(f)-h_2(f)|^2}{S(f)}df,
\end{aligned}
\ee
and $S(f)$ is the instrument noise spectrum.
In the comparison of the reflection model with the non-reflection one,
difference between two waveforms is given by $E_{ref}$ ($\widetilde{E}_{ref}$).
Here we test the distinguishability of the reflection signal using several analysis model for Cosmic Explorer and Einstein Telescope \cite{CE,ET,ET_D}.
In the following calculation, the distance $|CO| = 100\,\mbox{Mpc}$, $M_{\rm{BH}} = 10 M_{\odot}$, $M_{\rm{NS}} = 2 M_{\odot}$ and $r_{\rm NS} = 20 \,\rm km$.
An increasing trendency for $||\delta h||$ has been found as $f_{\rm GW}$ gets larger in the frequency region we are interested in.
The modulation signal is experimentally distinguishable in principle for $f_{\rm{GW}}$ which gives $||\delta h||>1$.
Due to the larger NS rotational radius in BNS system compared with BH in NSBH case,
more significant reflection signal should be expected from the former case,
which can be consistently observed from Fig. \ref{fig:SNR} for each type of detector model.
Furthermore, a wider observable region for the reflection signal is expected with more sensitive detectors in next generations.

\paragraph*{Conclusion.---}
\label{sec:conclusion}
To conclude,
we give an estimation to show the possible GW reflection by SF in the NSs,
given the GW frequency in observable region.
Based on this property,
we demonstrate the possibility of detecting SF in NSs based on GW measurements.
In both NSBH and BNS systems,
reflected GW signals contribute non-negligibly
to the total signal,
and modulate the amplitude and phase term at the order of
ratio of NS radius to orbital radius.
The modulation
is captured by the fluctuations of the amplitude and the linear phase,
which increases with the angular velocity.
Lastly, we show that such modulations caused by the GW reflection are detectable with Cosmic Explorer and Einstein Telescope in a reasonable parameter region.
The observation results after comparison with our theoretical
predictions
can provide evidence to support the existence of SF as
well as GW reflection of neutron stars.

Y. G., J. Y., and Z. Z. contribute equally to this study.

\bibstyle{apsrev-nourl}

\bibliography{Anti_GW}

\begin{thebibliography}{22}
\bibitem [{NSB()}]{NSBH_app}%
  \BibitemOpen
  \href@noop {} {}\bibinfo {note} {{The NS is initially set at the perigee,
  namely $\angle HCN(t) = \omega_{\alpha} t$. The distance relation is simply
  \be\ba &|ON(t)| = \sqrt{|CO|^2+R_{\rm NS}^2-2R_{\rm
  BH}|CH|\cos(\omega_{\alpha}t)}\\ &|N(t)B(t-\Delta t)|=\sqrt{R_{\rm
  NS}^2+R_{\rm BH}^2+2R_{\rm NS}R_{\rm BH}\cos(\omega_{\alpha}\Delta t)},
  \ea\ee with $R_{\rm BH} = (G(M_{\rm NS})^3/(M_{\rm NS}+M_{\rm
  BH})^2\omega_{\alpha}^2)^{1/3}$ and $R_{\rm NS} = (G(M_{\rm BH})^3/(M_{\rm
  NS}+M_{\rm BH})^2\omega_{\alpha}^2)^{1/3}$. GW signal received at $t_{obs}$
  is reflected by NS at a previous time $t_{ref}$, which comes from emission at
  time $t_{em}$. Subjecting to the GW propagation procedure we have \be\ba &
  c(t_{obs}-t_{ref}) = \sqrt{|CO|^2+R_{\rm NS}^2-2R_{\rm
  BH}|CH|\cos(\omega_{\alpha}t_{ref})}\\ & c \Delta t = \sqrt{R_{\rm
  NS}^2+R_{\rm BH}^2+2R_{\rm NS}R_{\rm BH}\cos(\omega_{\alpha}\Delta t)},
  \ea\ee with $\Delta t = t_{ref}-t_{em}$. By solving the set of equations
  numerically, we can find the propogation distance as well as the reflection
  amplitude and phase.}}\BibitemShut {Stop}%
\end{thebibliography}

\end{document}